\documentclass[journal]{IEEEtran}
\ifCLASSINFOpdf
\else
\fi
\usepackage{graphicx}      
\usepackage{cite}
\usepackage{amsmath}
\usepackage{amssymb}
\usepackage{epstopdf}
\newcommand{\mref}[1]{(\ref{#1})}

\begin{document}
\title{Comparison of frequency estimation methods for  wave energy control}

\author{Paula~B.~Garcia-Rosa,~\IEEEmembership{Member,~IEEE,}
        John~V.~Ringwood,~\IEEEmembership{Senior Member,~IEEE,}
        Olav~B.~Fosso,~\IEEEmembership{Senior Member,~IEEE,}
        and~Marta~Molinas,~\IEEEmembership{Member,~IEEE}
\thanks{Paula~B.~Garcia-Rosa and Marta~Molinas are with the Dept. of Eng. Cybernetics, Norwegian Univer. of Science and Technology, Trondheim, Norway (emails: p.b.garcia-rosa@ieee.org; marta.molinas@ntnu.no).}
\thanks{John~V.~Ringwood is with the Centre for Ocean Energy Research, Maynooth
University, Maynooth, Ireland (e-mail:john.ringwood@nuim.ie).}
\thanks{Olav~B.~Fosso is with the Dept.of Electric Power Eng., Norwegian Univ. of Science and Technology, Trondheim, Norway (email: olav.fosso@ntnu.no).}}
\maketitle

\begin{abstract}
A number of wave energy controllers tune the power take-off (PTO) system to the frequency of incident waves.  Since real ocean waves are non-stationary by nature and not defined by a single frequency component, the PTO can be either tuned at a constant frequency characterized by the local spectrum, or continuously tuned to a representative wave frequency. In either case, a time-frequency representation of the waves is expected since the wave profile changes over time.
This paper discusses about the PTO tuning problem for passive and reactive controllers, in real  waves, by comparing different methods for time-varying frequency estimation:  the extended Kalman filter (EKF), frequency-locked loop (FLL), and Hilbert-Huang transform (HHT).  The aim is to verify the impact of such methods on the absorbed and reactive powers, and the PTO rating.
It is shown that the mean estimated frequency of the EKF, and FLL, converges respectively to the mean centroid frequency, and  energy frequency, of the excitation force spectrum.  Moreover, the HHT mean frequency has no correlation with the spectral statistical properties.  A comparison of the energy absorbed shows that up to 37\% more energy is obtained with the HHT over the other estimation methods. Numerical  simulations are performed with sea elevation data from the Irish coast.


\end{abstract}

\begin{IEEEkeywords}
Energy harvesting, wave power, frequency estimation, control systems.
\end{IEEEkeywords}

%
\IEEEpeerreviewmaketitle


\section{Introduction}
\IEEEPARstart{I}{n} order to optimize the energy extracted from waves, a number of control algorithms for wave energy converters (WECs) tune the power take-off (PTO) system to the frequency of incident waves, see, e.g., \cite{Duclos2006,Yavuz2007,Korde2015,Cargo2016,Mendonca2016,Anderlini2017}.
For regular waves, the PTO tuning is well defined since the waves are characterized by a constant frequency sinusoidal motion. Nevertheless, real ocean waves are non-stationary by nature and not defined by a single frequency. Thus, the PTO can be either tuned to a constant representative frequency of the local wave spectrum, or continuously tuned to a  time-varying frequency.

The PTO tuning consists of adjusting the damping of the system, if a passive control method is adopted, or it may require adjustment of both the system damping and the stiffness for a reactive control method. The tuning strategies usually require knowledge of the incident wave frequency \cite{Duclos2006,Yavuz2007,Korde2015,Cargo2016,Mendonca2016} or the excitation force frequency \cite{Fusco2013,GRosa2017}.  Since these frequencies change with time for non-stationary signals, a time-frequency representation for the waves is expected -- i.e. the frequency is a function of time.

By adopting different methods to estimate the time-varying frequency for irregular waves and real ocean waves, some studies show that continuously tuning the PTO result in greater energy absorption than tuning the PTO to a constant frequency of the wave spectrum \cite{Yavuz2007,GRosa2017}. In \cite{Yavuz2007}, two methods have been adopted to estimate the on-line dominant wave frequency. One is based on the sliding discrete Fourier transform, while the other uses an analysis of the low-pass filtered incident wave, where wave characteristics, such as the zero-up crossing and crest-crest periods are calculated. In \cite{GRosa2017}, an estimation of the instantaneous frequency of the excitation force is obtained by means of the Hilbert-Huang transform (HHT), and such information is used for tuning the PTO damping of a WEC on a wave-by-wave basis. Other methods have adopted the extremum-seeking approach, where knowledge of the wave frequency is not needed for tuning purposes \cite{Hals2011,GRosa2012}. In such cases, the PTO parameters are adapted on an hourly basis (according to sea states variations) rather than a wave-by-wave basis.

Furthermore, a number of control algorithms also rely on the estimation of a time-varying frequency or the energy period of the waves \cite{Fusco2013,Anderlini2017,Fusco2014,Cantarellas2017,Tedeschi2012,Sakr2015}. An on-line estimate of the excitation force frequency is obtained with the extended Kalman filter (EKF) in \cite{Fusco2013}. In the proposed controller, the velocity reference is set as the ratio between the excitation force and the radiation damping, which is tuned to the estimated frequency \cite{Fusco2013}. An adaptive vectorial control approach is proposed in \cite{Cantarellas2017}, where a frequency-locked loop (FLL) is adopted to estimate the frequency of the WEC velocity.


The EKF-based method is based on a harmonic model with one variable frequency \cite{Fusco2013}. Then, the EKF tracks only a single dominant frequency. In contrast, the FLL method is based on an adaptive filter structure and an integral controller, where the selectivity of the filter and the tracking frequency dynamics can be adjusted \cite{Rodriguez2012}. Moreover, the instantaneous frequency, defined by the derivative of a phase function, is a time-varying parameter which identifies the location of the spectral peak of the signal as it varies with time \cite{Boashash1992}. However, the instantaneous frequency has physical meaning only for mono-component signals, \textit{i.e.}, signals with a single frequency or a narrow range of frequencies varying as a function of time \cite{Boashash1992}. In order to calculate the instantaneous frequency of multi-component signals, the HHT method \cite{Huang1998} firstly decomposes the original signal into a number of mono-component signals through the empirical mode decomposition (EMD). Such a decomposition has an adaptive basis and relies on the local characteristics of the signal. Thus, the EMD can extract different oscillation modes present in a wave profile.

The aim of this paper to verify the impact on the WEC performance of adopting different frequency estimation methods for tuning purposes in real ocean waves. Three methods are adopted for frequency estimation: EKF, FLL, and HHT. The methods are conceptually different and vary from estimating a single dominant frequency (EKF) to estimating the instantaneous wave-to-wave frequency of the oscillation modes present in a wave profile (HHT).

This study considers both passive and reactive controllers. For reactive control, the total power consists of a combination of active and reactive powers. Therefore, it is fundamental to individually determine these two power components to indicate the actual absorbed power, and the power that has to be supplied by the PTO during the conversion process. The performance of the WEC is then measured in terms of absorbed power, peak-to-average power ratio (PTO rating), and ratio of average reactive power and absorbed power. Numerical simulations with real sea elevation data from the Irish west coast are presented.


%
%
%
%

%
\section{Modeling and control of WECs}

\subsection{Equations of Motion}
This study considers a single oscillating-body represented as a truncated vertical cylinder constrained to move in heave.
%
%
%
%
With the assumption of linear hydrodynamic theory, and neglecting friction and viscous forces,
the motion of the floating cylinder is described by the superposition of the wave excitation force ($f_e$),
radiation force ($f_r$), restoring force ($f_s$) and the force produced by the PTO mechanism ($f_p$):
\begin{equation}
\label{eq:motion}
{m}{\ddot{x}}(t) = f_e(t) + f_r(t) + f_s(t) + f_p(t) \,,
\end{equation}
where ${x}\!\in\!\mathbb{R}$ is the vertical position of the body, and
$m\!\in\!\mathbb{R_+}$ is the body mass. The restoring force is given by $f_s=-Sx$,
where $S\!\in\!\mathbb{R_+}$ is the buoyancy stiffness.
%
%
From \cite{C:62}, the radiation force is calculated as 
\begin{equation}
  \label{eq:fr}
  - f_r(t)= m_{r}(\infty)\,\ddot{x}
            + \int\limits_{0}^{t}\!\!h_r(t-\tau)\,\dot{x}(\tau)\,d\tau\,,
\end{equation}
where ${m_r}(\infty)\!\in\!\mathbb{R_+}$ is the infinite-frequency added mass coefficient, defined with the asymptotic values of the added masses at infinite
frequency. The kernel of the convolution term ${h_r}(t-\tau)$ is known as the fluid memory term:
\begin{equation}
  \label{eq:Hr}
  h_r(t) = \frac{2}{\pi}\int\limits_{0}^{\infty}B_r(\omega)\cos(\omega t -\tau)\,d\omega\,, \\
\end{equation}
where $B_{r}(\omega)\!\in\!\mathbb{R_+}$ is the radiation damping coefficient, and $\omega\!\in\!\mathbb{R_+}$ is the wave frequency.
%
%
From \mref{eq:motion} and \mref{eq:fr}, 
\begin{eqnarray}
 \label{eq:x_fefp}
M\ddot{x}(t)\!+\!\int\limits_{0}^{t}\!\!h_r(t-\tau)\,\dot{x}(\tau)\,d\tau
\!+\! Sx(t) \!=\!  f_e(t)\!+\!f_p(t)\,,
\end{eqnarray}
with $M\!=\![m + m_{r}(\infty)]$. The excitation force, i.e., the force due to the incident waves is given by
\begin{eqnarray}
  \label{eq:fet}
  f_e(t)= \int\limits_{-\infty}^{\infty}\!\!h_e(t-\tau)\,\zeta(\tau)\,d\tau\,, \\
  \label{eq:he}
  h_e(t) = \frac{1}{2\pi}\int\limits_{-\infty}^{\infty}H_e(\omega)e^{i\omega t}\,d\omega\,.
\end{eqnarray}
%
%
$h_e$ is the inverse Fourier transform of the excitation force transfer function ${H_e}(\omega)$, which has low-pass filter characteristics for floating WECs, and $\zeta$ is the wave elevation. Notice that \mref{eq:he} is non-causal, since in fact, the pressure distribution is the cause of the force and not the incident waves \cite{F:02}.

\subsection{Passive Control}

A generic PTO mechanism, with a damper varying in time ($B_p\!\in\!\mathbb{R_+}$), is considered for the passive control (PC). Thus,
\begin{equation}
\label{eq:fp_p}
f_p(t)=-B_p(t)\dot{x}(t)\,.
\end{equation}

For the case when $B_p(t)\!=\!B_p$, for any time $t$, and for monochromatic waves, the maximum absorption is obtained when \cite{F:02}:
\begin{equation}
\label{eq:Bp}
B_p=\sqrt{(B_r(\omega))^2 + (\omega(m+m_r(\omega))-S/\omega)^2}\,,
\end{equation}
where $m_r(\omega)\!\in\!\mathbb{R}$ is the added mass. Equation \mref{eq:Bp} is frequency dependent, and indicates that there is an optimal damping
for each frequency when the WEC is submitted to real ocean waves, or irregular waves with a mixture of frequencies.


Here, the PTO damping is continuously modified, and tuned to the excitation force frequency. From \mref{eq:Bp},
\begin{equation}
\label{eq:Bt}
B_p(t)=\sqrt{(B_r(\hat\omega))^2 + (\hat\omega(m+m_r(\hat\omega))-S/\hat\omega)^2}\,,
\end{equation}
where $\hat\omega(t)$ is the estimated time-domain frequency of the wave excitation force. In order to examine the impact of the time-varying frequency on the performance of the WEC, three methods are adopted for the frequency estimation: EKF, FLL, and HHT.

\subsection{Reactive Control}
For reactive control (RC), the PTO force consists of a damping term and a spring term:
\begin{equation}
\label{eq:fp_r}
f_p(t)=-B_p(t)\dot{x}(t) - S_p(t)x(t)\,,
\end{equation}
where $S_p\!\in\!\mathbb{R}$ is the stiffness coefficient. For incident regular waves, if $S_p(t)\!=\!S_p$ for any time $t$, and
\begin{equation}
\label{eq:Sp}
S_p=\omega^2(m+m_r(\omega))-S\,,
\end{equation}
then the reactive part of the total impedance $$[R_p+R_r(\omega)]+j\omega[(m+m_r(\omega)-(S+S_p)/\omega^2)]$$ is cancelled and the velocity of the floating body is in phase with the excitation force \cite{F:02}. In such a case, the PTO damping \mref{eq:Bp} becomes
\begin{equation}
\label{eq:BpRC}
B_p=B_r(\omega)\,,
\end{equation}
and the greatest wave energy absorption is obtained.

Following the same procedure for PC, $S_p$ and $B_p$ are tuned to the excitation force frequency. From \mref{eq:Sp} and \mref{eq:BpRC},
\begin{eqnarray}
\label{eq:SpRpRC}
S_p(t)&\!\!=\!\!&\hat\omega^2(t)(m+m_r(\hat\omega(t)))-S \,,\\
B_p(t)&\!\!=\!\!& B_r(\hat\omega(t))\,.
\end{eqnarray}

Notice that, for practical application studies, the physical limits of the body excursion and the PTO should be taken into account. Here, a PTO force constraint is implemented as a saturation. However, this is only a theoretical approach, since the body motion is also a function of the excitation force, which is an external force that cannot be manipulated. The implementation of saturation on the force signals of a real WEC is not physically possible.

\subsection{Energy and Power}

The mean power and energy absorbed by the WEC over a time range $T$ are, respectively,
\begin{equation}
\label{eq:E}
\bar{P}=\frac{E}{T}\,, \quad \text{and} \quad
E=-\int_{0}^{T}B_{p}(t)\dot{x}^2(t)dt\,.
\end{equation}

Notice that, for the reactive control, the delivered power has two components: the absorbed power (or active power, that is the power delivered to the damping $B_p$), and the reactive power (the power delivered to the spring $S_p$) \cite{F:02}. The mean reactive power, and reactive energy, are respectively,
\begin{equation}
\label{eq:Er}
\bar{P}_r=\frac{E_r}{T}\,, \quad \text{and}\quad
E_r=-\int_{0}^{T}S_{p}(t)x(t)\dot{x}(t)dt\,.
\end{equation}
%
%
Since the spring term in the PTO force \mref{eq:fp_r} and the body velocity can have opposite signs in \mref{eq:Er}, the PTO has to return energy for some parts of the wave cycle. Then, the PTO system should be able to implement bidirectional power flow for RC.

%

In this study, the performance of the WEC in terms of mean absorbed power ($\bar{P}$) is measured by the capture width ratio,
\begin{equation}
\text{CWR} = \dfrac{\bar{P}}{2rP_\zeta}\,,
\end{equation}
where $r$ is the cylinder radius and $P_\zeta$ is the transported wave power per unit width of the wave front. In deep water \cite{F:02},
\begin{equation}
P_\zeta=\dfrac{\rho g}{2}\int_0^\infty \dfrac{S_\zeta(\omega)}{\omega}d\omega\,,
\end{equation}
where $S_\zeta$ is the wave spectrum, $\rho$ is the sea water density, and $g$ is the gravitational acceleration.






\section{Estimation of the wave excitation force frequency}

\subsection{Extended Kalman Filter}

In order to estimate the frequency by means of the EKF, we assume a harmonic model with a single time-varying frequency component and amplitude, as proposed in \cite{Fusco2010}. Thus, $f_e(t)$ can be expressed in discrete-time ($t=kT_s$) as:
\begin{equation}
\label{eq:fe_ekf}
f_{e}[k] = A[k] \cos{(\omega[k] kT_s + \varphi[k]) + \eta[k]}\,,
\end{equation}
where $A$ is the amplitude of the wave excitation force, $\varphi$ is the phase, and $T_s$ is the sampling time. Following the cyclical structural model from \cite{Harvey1989}, \mref{eq:fe_ekf} can be written in a recursive non-linear state space form as
\begin{equation}
\begin{aligned}
\label{eq:m_ekf}
\upsilon[k+1]&=f(\upsilon[k]) + \varpi[k] \\
f_e[k]&=h(\upsilon[k]) + \eta[k]
\end{aligned}
\end{equation}
where $\upsilon\!\!\in\!\!\mathbb{R}^{3}$ is the state vector defined as $\upsilon\!\!=\!\!\begin{bmatrix}\psi&\psi^{*}&\omega\end{bmatrix}^T$,
${\varpi\!\in\!\mathbb{R}^{3}}$ and ${\eta\!\in\!\mathbb{R}}$ are zero-mean independent random processes with covariance matrices defined, respectively, as
$E[\varpi\varpi^T]\!=\!R$ and $E[\eta\eta^T]\!=\!Q$. Functions $f(\cdot)$ and $h(\cdot)$ are,
$$f(\upsilon[k])=\begin{bmatrix}
\cos{(\omega[k] T_sk)} & \sin{(\omega[k] T_sk)} & 0 \\
-\sin{(\omega[k] T_sk)}& \cos{(\omega[k] T_sk)} & 0 \\
0 & 0 & 1
\end{bmatrix}\upsilon[k],$$
and
$h(\upsilon[k])\!=\!\begin{bmatrix}
1&0&0\end{bmatrix}\upsilon[k]$, respectively \cite{Fusco2010}.

The EKF obtains an estimate of the state vector $\upsilon[k]$ based on observations of $f_e[k]$, and on the first-order linearization of model \mref{eq:m_ekf} around the last state estimate. The EKF algorithm is summarized in Table \ref{tab:EKF}, where $I$ is the identity matrix of order 3, $J_f$ and $J_h$ are the Jacobian matrices of $f(.)$ and $h(.)$, denoted respectively by $J_f[k]=\nabla f |_{\hat\upsilon[k|k]}$ and $J_h[k+1]=\nabla h|_{\hat\upsilon[k+1|k]}$.
\begin{table}[!htpb]
\begin{center}
  \caption{EKF algorithm.}
\begin{tabular}{|l|}
  \hline
  Prediction step:  \\
  $\hat\upsilon[k + 1|k]= f(\hat\upsilon[k|k])$ \\
  $P[k + 1|k] = J_f[k]P[k|k]J_f[k]^T + Q[k]$ \\ \hline
  Innovation step: \\
  $\hat{\upsilon}[k + 1|k + 1] = \hat{\upsilon}[k + 1|k]$ \\
  \hspace{2.3cm}$ + K[k+1](f_e[k+1]-h(\hat\upsilon[k + 1|k]))$\\
  $K[k + 1] = P(k + 1|k)J_h[k+1]^T $ \\
  \hspace{2.3cm}$(J_h[k+1]P[k + 1|k]J_h[k+1]^T+ R[k + 1])^{-1}$ \\
  $P[k + 1|k + 1] = (I -K[k + 1]J_h[k+1])P[k + 1|k]$ \\ \hline
\end{tabular}
 \label{tab:EKF}
\end{center}
\end{table}

Once an estimate of the state vector $\hat\upsilon[k|k]$ is available from the EKF (Table \ref{tab:EKF}), the amplitude and frequency of the wave excitation force are, respectively, obtained as:
\begin{eqnarray}
\hat{A}_{_\text{EKF}}[k|k] &=&\sqrt{\hat\psi[k|k]^2 + \hat\psi^{*}[k|k]^2}\,, \\
\hat\omega_{_\text{EKF}}[k|k] &=& \hat\omega[k|k]\,.
\end{eqnarray}
%

\subsection{Frequency-Locked Loop}
\begin{figure}[!htpb]
\begin{center}
  \includegraphics[width=8cm]{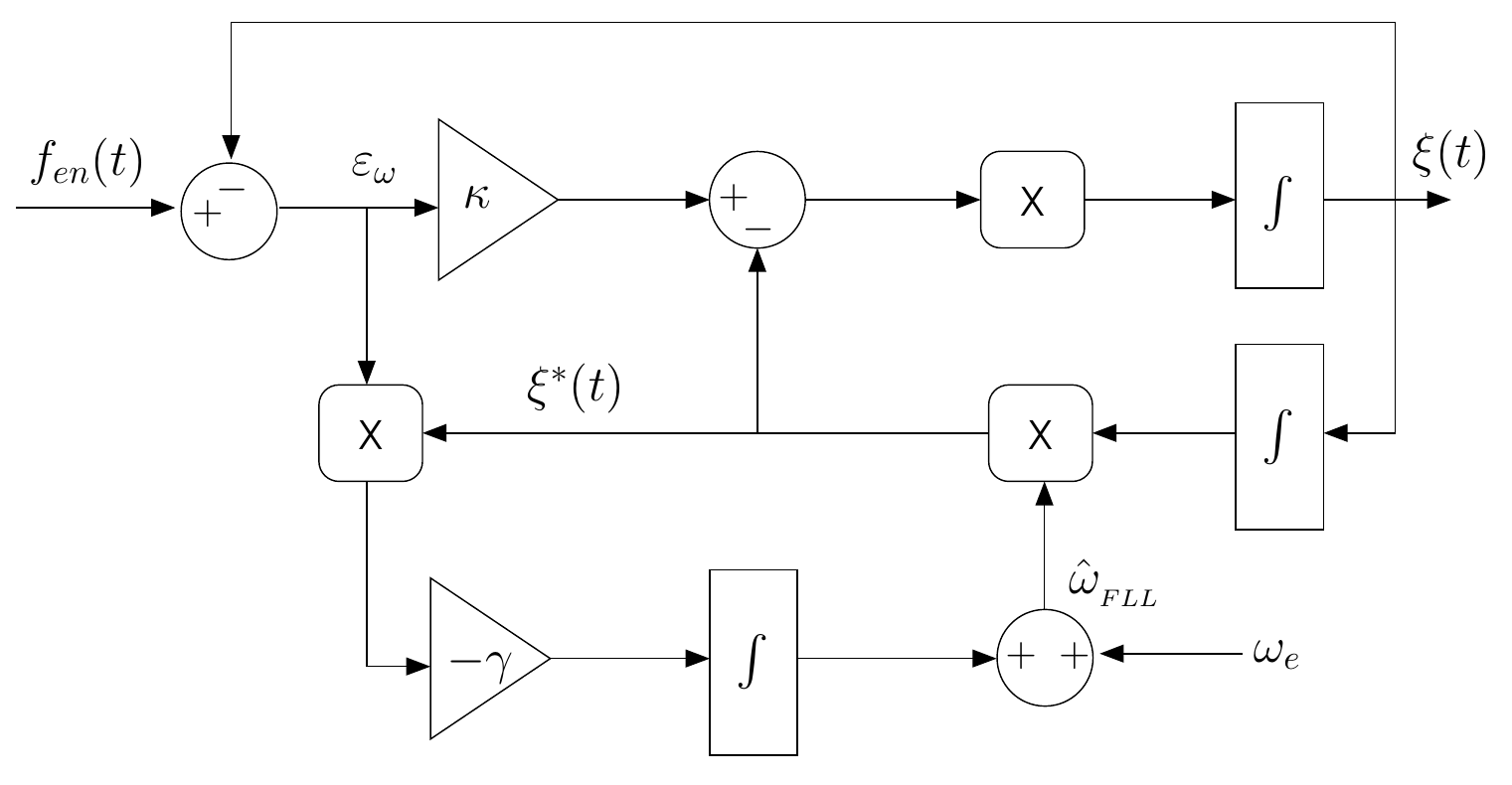}
\end{center}
  \caption{Frequency estimation by the SOGI-FLL method.}
\label{fig:FLL}
\end{figure}

An adaptive filter is implemented by means of a second-order generalized integrator (SOGI), where the FLL estimates the frequency of the input signal. Such a scheme is termed SOGI-FLL, and has been proposed for grid synchronization of power converters \cite{Rodriguez2012}.
Figure \ref{fig:FLL} illustrates the block diagram of the SOGI-FLL, consisting of the SOGI-QSG (SOGI quadrature signal generator) and the FLL structure. The method was first adopted, within the wave energy control context, to estimate the frequency components of the WEC velocity \cite{Cantarellas2017}. In this study, the method is adopted for estimating the wave excitation force frequency.

The SOGI-QSG acts as an adaptive bandpass filter with two in-quadrature output signals ($\xi$, $\xi^{*}$), where $\xi^{*}$ lags $\xi$ by $90^\circ$. The bandwidth of the filter is exclusively set by the gain $\kappa$ \cite{Rodriguez2012}. The FLL is responsible for estimating the frequency of the input signal. Notice that the in the nonlinear frequency adaptation loop, $\varepsilon_\omega\!=\!\varepsilon \xi^{*}$ can be interpreted as a frequency error variable, and the parameter $\gamma$ represents the gain of the integral controller \cite{Rodriguez2012}. The selectivity of the adaptive bandpass filter and the tracking frequency dynamics are respectively determined by the tuning parameters $\kappa$ and $\gamma$.


From Fig.\,\ref{fig:FLL}, the state-space equations of the SOGI-FLL are
\begin{eqnarray}
\label{eq:sogi}
\begin{bmatrix}
\dot\xi \\
\dot\nu \\
\end{bmatrix}&\!\!=\!\!&\begin{bmatrix}
-\kappa\hat\omega_{_\text{FLL}} & -\hat\omega_{_\text{FLL}}^2  \\
0 & 1 \\
\end{bmatrix}\,
\begin{bmatrix}
\xi \\
\nu \\
\end{bmatrix}\,
+
\begin{bmatrix}
\kappa\hat\omega_{_\text{FLL}} \\
0\\
\end{bmatrix}
f_{en} \nonumber \\
\begin{bmatrix}
\xi \\
\xi^{*} \\
\end{bmatrix}&\!\!=\!\!&\begin{bmatrix}
1 & 0  \\
0 & \hat\omega_{_\text{FLL}} \\
\end{bmatrix}\,
\begin{bmatrix}
\xi \\
\nu \\
\end{bmatrix}\, \\
%
%
\label{eq:fll}
\dot{\hat{\omega}}_{_\text{FLL}}&\!\!=\!\!&-\gamma(f_{en}-\xi)\xi^{*}\,\,\hat\omega_{_\text{FLL}}\,,
%
\end{eqnarray}
where $(\xi,\,\nu)$ and $(\xi,\,\xi^{*})$ are, respectively, the state and output vectors of the SOGI, and $f_{en}$ is the normalized excitation force. The FLL state equation is represented by \mref{eq:fll}.

\subsection{Hilbert-Huang Transform}
\begin{figure}[!htpb]
\begin{center}
  \includegraphics[width=8cm]{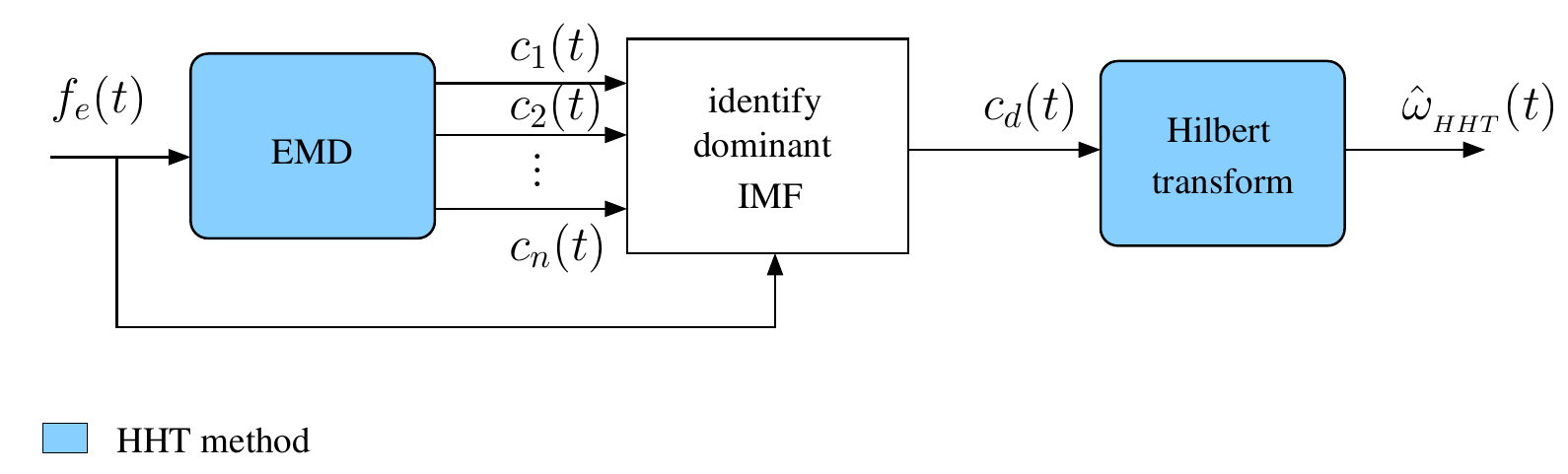}
\end{center}
  \caption{Frequency estimation by the HHT method.}
\label{fig:HHT}
\end{figure}

In the HHT method, the wave excitation force $f_e(t)$ is firstly decomposed into $N$ mono-component signals (IMFs) by the EMD. Then, the instantaneous frequency of the dominant IMF is adopted for tuning purposes \cite{GRosa2017}. Figure \ref{fig:HHT} illustrates the block diagram of the frequency estimation by this method.

The EMD identifies local maxima and minima of $f_e(t)$, and calculates upper and lower envelopes for such extrema using cubic splines. The mean values of the envelopes are used to decompose the original signal into frequency components in a sequence from the highest frequency to the lowest one.
The EMD procedure is summarized in Table \ref{tab:EMD}.

\begin{table}[!htpb]
\begin{center}
  \caption{EMD algorithm.}
\begin{tabular}{|l|}
  \hline
Step 0: Set $i\!=\!1$; $r(t)\!=\!f_e(t)$; \\ \hline
Step 1: Identify the local maxima and minima in $r(t)$;\\ \hline
Step 2: Calculate the upper envelope defined by the maxima, \\ \hspace{1cm}and the lower envelope defined by the minima; \\\hline
Step 3: Calculate the mean envelope $m(t)$; \\ \hline
Step 4: Set $h(t)\!=\!r(t)-m(t)$; \\ \hline
Step 5: If $h(t)$ is an IMF, go to next step. Otherwise, set $r(t)\!=\!h(t)$ \\ \hspace{1cm} and  go back to step 1; \\ \hline
Step 6: Set $c_i(t)\!=\!h(t)$; $r(t)\!=\!r(t)-c_i(t)$; \\ \hline
Step 7: If $i\!=\!N$, define the IMF components as $c_1(t),\ldots,c_N(t)$, \\
\hspace{1cm} and the residue as $r(t)$. Otherwise, set $i\!=\!i+1$ and \\ \hspace{1cm} go back to step 1. \\ \hline
\end{tabular}
 \label{tab:EMD}
\end{center}
\end{table}

Then, the wave excitation force can be expressed as
\begin{equation}
\label{eq:emd}
    f_e(t) =\sum_{i=1}^{N} c_{i}(t) + r(t)\,,
\end{equation}
where $N$ is the total number of IMFs, which is defined here as $\log_2N_s\!-\!1$ \cite{Wu2004}, $N_s$ is the data length, and $r(t)$ is the residue.

The dominant IMF is identified through the comparison of the energy of the IMF signals ($E_{c_i}$) with the energy of the excitation force signal ($E_{f_e}$),
\begin{equation}
E_{c_i}=\int_{0}^{T}|{c}_i(t)|^2 dt\,,  \quad
E_{f_e}=\int_{0}^{T}|f_e(t)|^2 dt\,,
\end{equation}
where $c_i(t)$ is the $i$-th IMF component. The dominant component $c_d(t)$ is the IMF with the highest $E_{c_i}/E_{f_e}$ ratio.

%
Finally, the Hilbert transform (HT) is applied to ${c}_d(t)$ \cite{Huang1998}:
\begin{equation}
\label{eq:HT}
\upsilon_{d}(t)=\frac{1}{\pi}\,P\,\int_{-\infty}^{\infty}\frac{{c}_d(\tau)}{t-\tau}d\tau\,,
\end{equation}
where $P$ indicates the Cauchy principal value. Then, the dominant IMF is represented as an analytic signal,
%
\begin{equation}
\label{eq:zi}
{z}_d(t) = {c}_d(t) + j\upsilon_{d}(t)\,,
\end{equation}
with amplitude $\hat{A}_{_\text{HHT}}$, phase $\hat\phi_{_\text{HHT}}$, and instantaneous frequency $\hat\omega_{_\text{HHT}}$, respectively estimated as
%
\begin{eqnarray}
\label{eq:at}
\hat{A}_{_\text{HHT}}(t)=\sqrt{{c}_d^2(t)+\upsilon_d^2(t)}\,,
\hat\phi_{_\text{HHT}}(t)=\arctan\left(\dfrac{\upsilon_d(t)}{{c}_d(t)}\right)\,,
\end{eqnarray}
\begin{equation}
\label{eq:wt_hht}
\hat\omega_{_\text{HHT}}(t)=\dot\phi_d(t)\,.
\end{equation}
%


\section{Simulation results}

\subsection{Hydrodynamic parameters}

The same heaving cylinder adopted in \cite{GRosa2017} is considered here. The cylinder has a radius of $r\!=\!5$\,m, draught $d\!=\!4$\,m, mass $m\!=\!3.2\times10^5$\,kg and resonance frequency $1.2$\,rad/s. The hydrodynamic coefficients of the cylinder were computed using the boundary element solver WAMIT \cite{W:98}. The added mass, radiation damping coefficients, and the frequency response of the excitation force are shown in \cite{GRosa2017}.
%


\subsection{Real sea elevation data}

The wave data was collected in 2010 from a data buoy in the Belmullet wave energy test site, off the west coast of Ireland. The wave data, provided by the Irish Marine Institute, consists of wave elevation records of $30$ minutes sampled at $1.28$\,Hz.

\begin{figure}[!htpb]
\begin{center}
  \includegraphics[width=8cm]{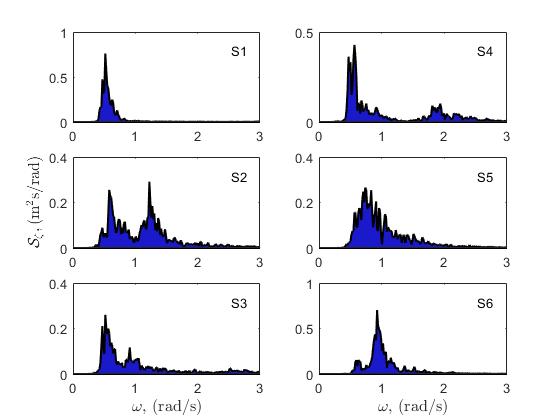}
\end{center}
  \caption{Wave spectra of real wave data from Belmullet.}
\label{fig:wave_spectra}
\end{figure}

Six wave elevation records (referred as sea states S1-S6), with different spectral distribution, were selected for our study. Figure \ref{fig:wave_spectra} illustrates the wave spectra of the sea states, and Table \ref{tab:seastates} shows the significant wave height ($H_s$), the peak frequency ($\omega_p$), the energy frequency ($\omega_e$), and the mean centroid frequency ($\omega_1$) of the spectra.
The statistical parameters $H_s$, $\omega_e$ and $\omega_1$ are respectively calculated as: $H_s\!=\!4\sqrt{m_0}$, $\omega_e\!=\!m_0/m_{-1}$, $\omega_1\!=\!m_0/m_1$, where $m_n\!=\!\int_{0}^{\infty}\omega^nS(\omega)d\omega$ is the spectral moment of order $n$. $\omega_p$ is the frequency at which the wave spectrum is maximum.

\begin{figure}[!htpb]
\begin{center}
  \includegraphics[width=8cm]{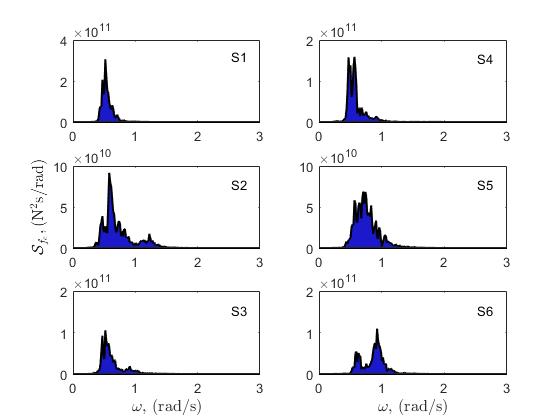}
\end{center}
  \caption{Excitation force spectra for sea states S1-S6.}
\label{fig:force_spectra}
\end{figure}

Figure \ref{fig:force_spectra} shows the spectral density of the excitation force for the selected sea states. Some of the high frequency waves are filtered out by the transfer function $H_e(\omega)$, as can be noted from Figure \ref{fig:wave_spectra}. The filtering characteristics are defined by the shape of the
floating body, so that the excitation force spectra are characteristic of the cylinder adopted in this study.
\begin{table}[!htpb]
 \begin{center}
  \caption{Significant wave height $H_s$\,(m), peak frequency $\omega_p$\,(rad/s),
  energy frequency $\omega_e$\,(rad/s), and mean centroid frequency $\omega_1$\,(rad/s) of the selected sea states.}
  \begin{tabular}{|c|c|c|c|c|c|c|}
  \hline
           &S1 & S2 & S3 & S4 & S5 & S6  \\ \hline
 $H_s$     &\!1.26\!&\!1.43\!&\!1.18\!&\!1.39\!&\!1.42\!&\!1.62 \\ \hline
 $\omega_p$&\!0.52\!&\!1.22\!&\!0.52\!&\!0.57\!&\!0.74\!&\!0.93 \\ \hline
 $\omega_e$&\!0.59\!&\!0.94\!&\!0.80\!&\!0.80\!&\!0.90\!&\!0.97 \\ \hline
 $\omega_1$&\!0.70\!&\!1.18\!&\!1.13\!&\!1.18\!&\!1.06\!&\!1.08 \\ \hline
  \end{tabular}
 \label{tab:seastates}
 \end{center}
\end{table}

\subsection{Time-frequency estimation by EKF, FLL and HHT}

\subsubsection{Superposition of two regular waves} In order to illustrate how the estimated excitation force frequency differs according to the method adopted, firstly we consider a simple incident wave defined as the superposition of two regular waves: $\zeta(t)\!=\!2\cos{(2\pi/6t)}+\cos{(2\pi/8t)}$. The energy frequency and the mean centroid frequency of the excitation force spectra are, respectively, $\omega_{e,f_e}\!=\!0.94$\,rad/s and $\omega_{1,f_e}\!=\!0.96$\,rad/s.

\begin{figure}[!htpb]
\begin{center}
  \includegraphics[width=7.5cm]{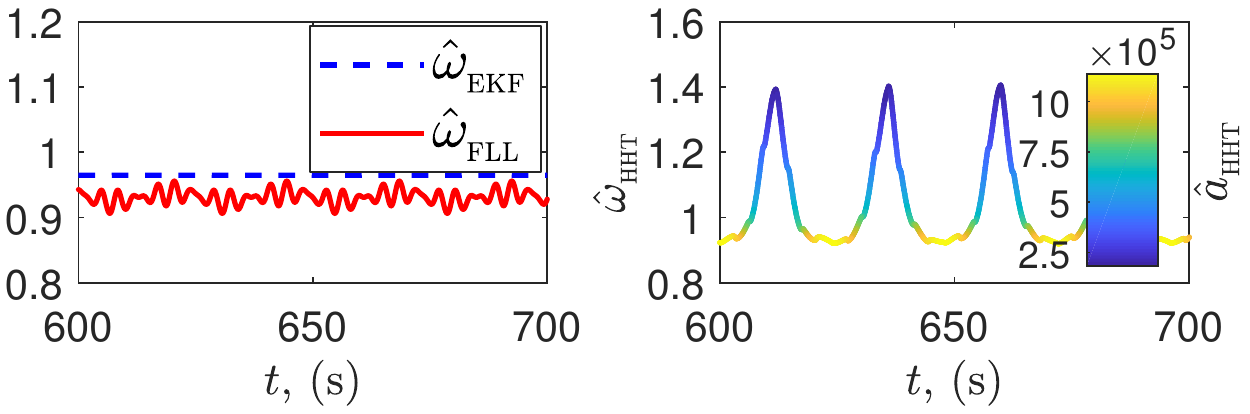}
\end{center}
  \caption{Frequency estimated (rad/s) by the EKF, FLL (left) and HT (right) for the wave $\zeta(t)\!=\!2\cos{(2\pi/6t)}+\cos{(2\pi/8t)}$. The plot on the right represents the Hilbert spectrum.}
\label{fig:wt_irreg}
\end{figure}

Figure \ref{fig:wt_irreg} illustrates the excitation force frequency estimated by the EKF, FLL, and HT. The Hilbert spectrum shows that the instantaneous frequency varies from about 0.92 to 1.4 rad/s, with the highest energy content (highest amplitude) in the lowest frequency. The EKF tracks a single frequency (0.96 rad/s) which represents the mean centroid frequency of the excitation force spectrum, and the mean frequency estimated by the FLL (0.93\,rad/s) is close to the energy frequency of the spectrum. The frequency estimated by the EKF is nearly constant, while the tracking frequency dynamics of the FLL depends mainly on the selection of the parameters $\kappa$ and $\gamma$. To ensure high frequency selectivity, and accurate direct and quadrature components within frequency range 0.6 to 1\,rad/s, $\kappa$ is set to $\sqrt{2}$ and $\gamma\!=\!0.16$, as discussed in \cite{Cantarellas2017}.

\subsubsection{Sea Elevation Data}
\label{sec:realdata}
\begin{figure*}[!t]
\begin{center}
  \includegraphics[width=7.5cm]{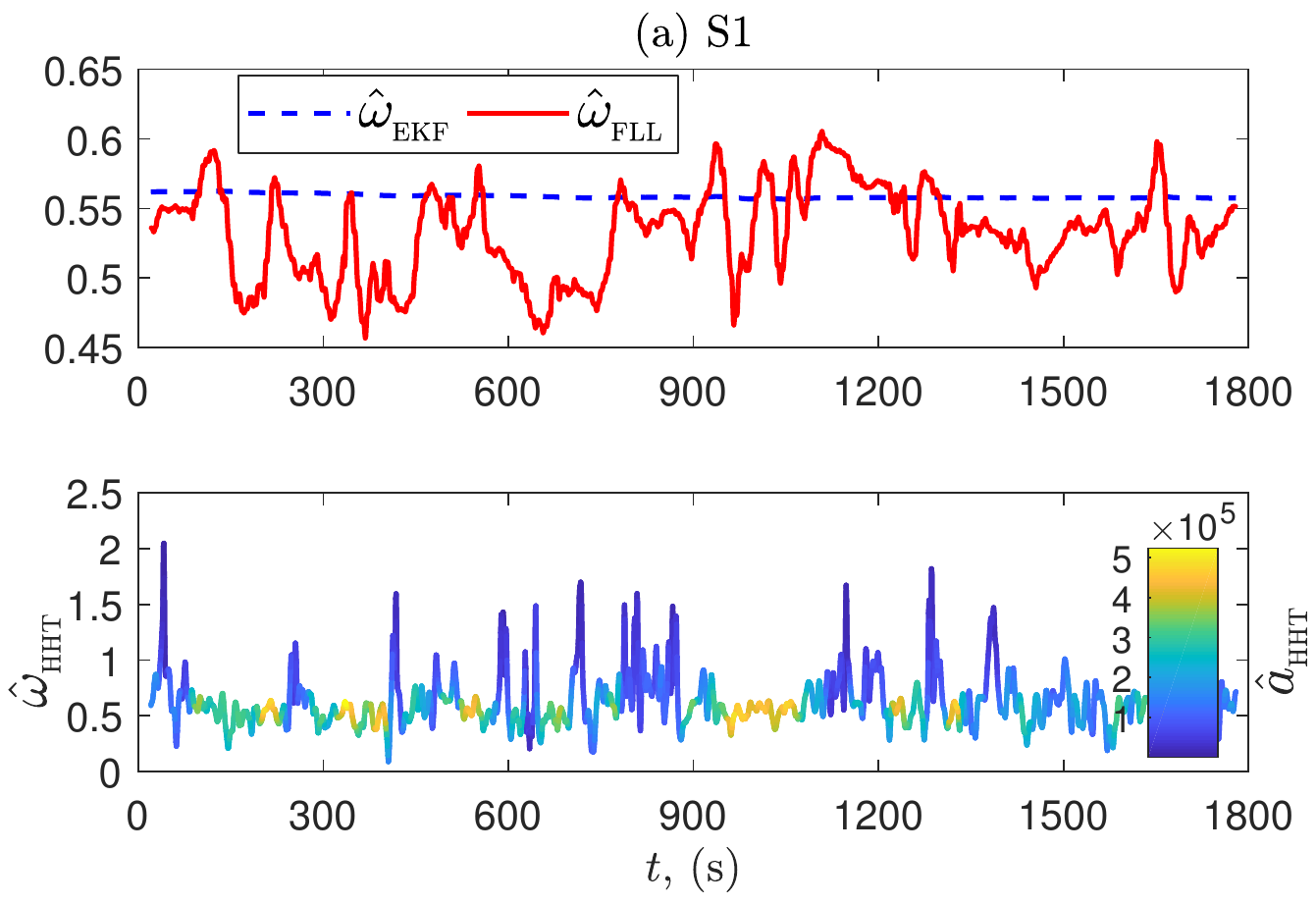}
  \includegraphics[width=7.5cm]{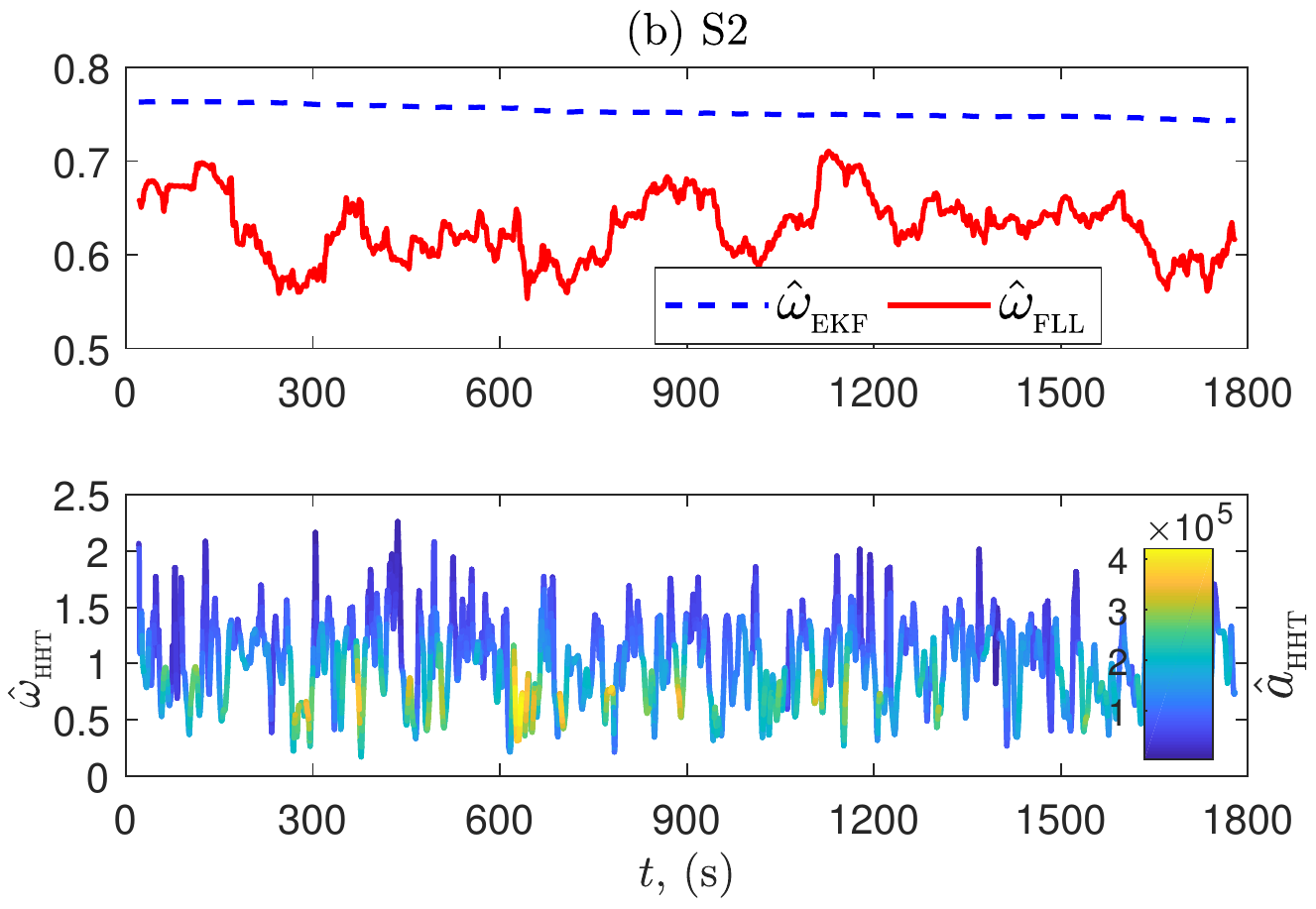}
\end{center}
  \caption{Frequency estimated (rad/s) by the EKF, FLLt and HHT for (a) S1 and (b) S2. The plots in the bottom represent the Hilbert spectrum of the first IMF.}
\label{fig:wt_S1_S2}
\end{figure*}
Table \ref{tab:wforcespectra} shows the energy and mean centroid frequencies of the excitation force spectra, and the mean frequencies estimated by the studied methods for sea states S1 to S6. It can be noted that the mean values of the frequencies estimated by the EKF converge to the mean centroid frequency of the spectra, while the mean frequency values of the FLL converge to the energy frequency of the spectra. However, the mean values estimated by the HHT have no correlation with the statistical parameters obtained from the spectra. As has been remarked by Huang et al. \cite{Huang1999}, frequency in the Hilbert spectrum has a different meaning from Fourier spectral analysis. In Fourier spectral analysis, the existence of energy at a frequency means that a component of a sine (or a cosine) wave persisted through the entire time range of the data, whereas in the Hilbert spectrum the wave representation is local and the exact time of such oscillation is given \cite{Huang1999}.
\begin{table}[!htpb]
 \begin{center}
  \caption{Energy frequency ($\omega_e$), mean centroid frequency ($\omega_1$) of the excitation force spectra, and mean values of the estimated frequencies
  ($\bar{\omega}_{_\text{EKF}}$, $\bar{\omega}_{_\text{FLL}}$, $\bar{\omega}_{_\text{HHT}}$). Frequencies in rad/s.  }
  \begin{tabular}{|c|c|c|c|c|c|c|}
  \hline
           &S1 & S2 & S3 & S4 & S5 & S6  \\ \hline
 $\omega_{e,f_e}$&\!0.54\!&\!0.66\!&\!0.59\!&\!0.56\!&\!0.72\!&\!0.82 \\ \hline
 $\omega_{1,f_e}$&\!0.55\!&\!0.73\!&\!0.63\!&\!0.59\!&\!0.76\!&\!0.87 \\ \hline\hline
 $\bar{\omega}_{_\text{EKF}}$&\!0.56\!&\!0.75\!&\!0.65\!&\!0.60\!&\!0.78\!&\!0.88 \\ \hline
 $\bar{\omega}_{_\text{FLL}}$&\!0.53\!&\!0.63\!&\!0.57\!&\!0.55\!&\!0.70\!&\!0.80 \\ \hline
 $\bar{\omega}_{_\text{HHT}}$&\!0.62\!&\!1.01\!&\!0.80\!&\!0.73\!&\!0.90\!&\!0.97 \\ \hline
  \end{tabular}
 \label{tab:wforcespectra}
 \end{center}
\end{table}

The estimated frequencies for sea states S1 and S2 are illustrated in Figure \ref{fig:wt_S1_S2}. From the Hilbert spectrum of the first IMF, it can be noted that the frequency ranges $0.5-0.6$\,rad/s, and $0.6-0.8$\,rad/s, have the highest energy content, respectively, for S1 and S2. Such frequency values coincide with the estimates from the EKF and FLL, but the HHT method identifies the time at which such oscillations occur. Notice that the HHT analysis returns nine IMFs, and the first IMF is the dominant component in all studied cases, as shown in \cite{GRosa2017}.

%

\subsection {Effect of the estimated frequency on the control strategy}
In order to limit the body excursions to $2.5$\,m for the studied cases, the PTO force of the PC \mref{eq:fp_p} and each term of the RC in \mref{eq:fp_r} is limited to $\pm500$\,kN.
%

\subsubsection{Passive Control}
\begin{figure}[!htpb]
\begin{center}
  \includegraphics[width=7.5cm]{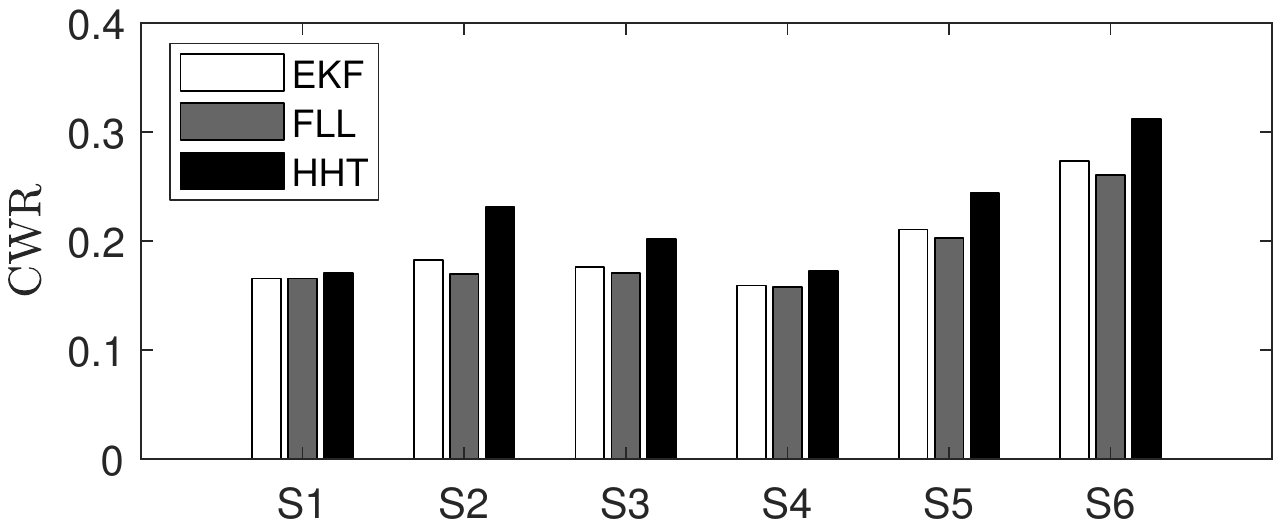}
\end{center}
  \caption{CWR under PC tuned at frequencies from the EKF, FLL and HHT.} 
\label{fig:cwr_pc}
\end{figure}

\begin{figure}[!htpb]
\begin{center}
  \includegraphics[width=7.5cm]{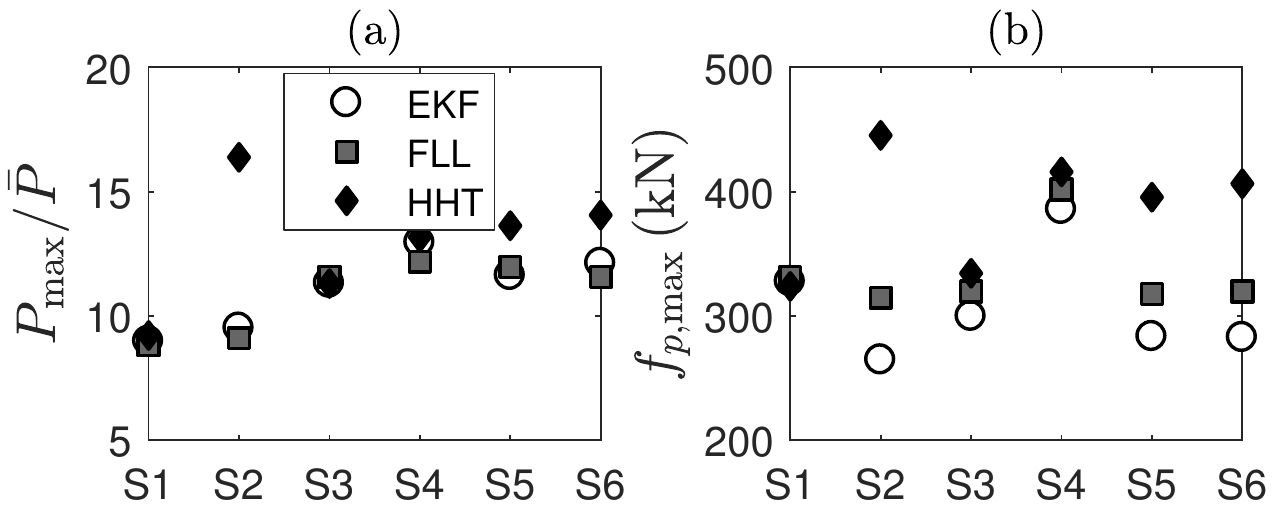}
\end{center}
  \caption{WEC performance under PC (a) Peak-to-average power ratio (b) Maximum PTO force required.}
\label{fig:p2a_pc}
\end{figure}

The performance of the WEC is illustrated in Figures \ref{fig:cwr_pc} and \ref{fig:p2a_pc}, for the cases when the PC strategy adopts the EKF, FLL or HHT methods to estimate the excitation force frequency for sea states S1 to S6 (section\,\ref{sec:realdata}).

For all the studied cases, tuning the damping with frequency estimates from the HHT gives greater energy capture than tuning with EKF and FLL. The highest improvement is of a factor of $1.27$ when HHT is compared to EKF, or $1.37$ when compared to FLL, for sea state S2. Moreover, the lowest differences in the CWR is obtained for S1. Such behaviour can be explained by the different energy spectral distributions of both sea states. S1 is characterized by a narrowband spectrum with a single dominant swell (low frequency waves generated in other locations), with the energy concentrated in a narrow band of frequencies. However, S2 is characterized by a two-peak spectrum with mixed wind-sea (high frequency waves generated by the local wind) and swell conditions, with the energy spread over a wider band of frequencies than S1. In such a case, a method that calculates the wave-to-wave frequency is more beneficial for PC than a method that gives a dominant sea state frequency. Nevertheless, the PTO rating and the maximum PTO required for the HHT frequencies are also higher than for the EKF and FLL, especially for S2 (Fig.\,\ref{fig:p2a_pc}).

%

\subsubsection{Reactive Control}
\begin{figure}[!htpb]
\begin{center}
  \includegraphics[width=7.5cm]{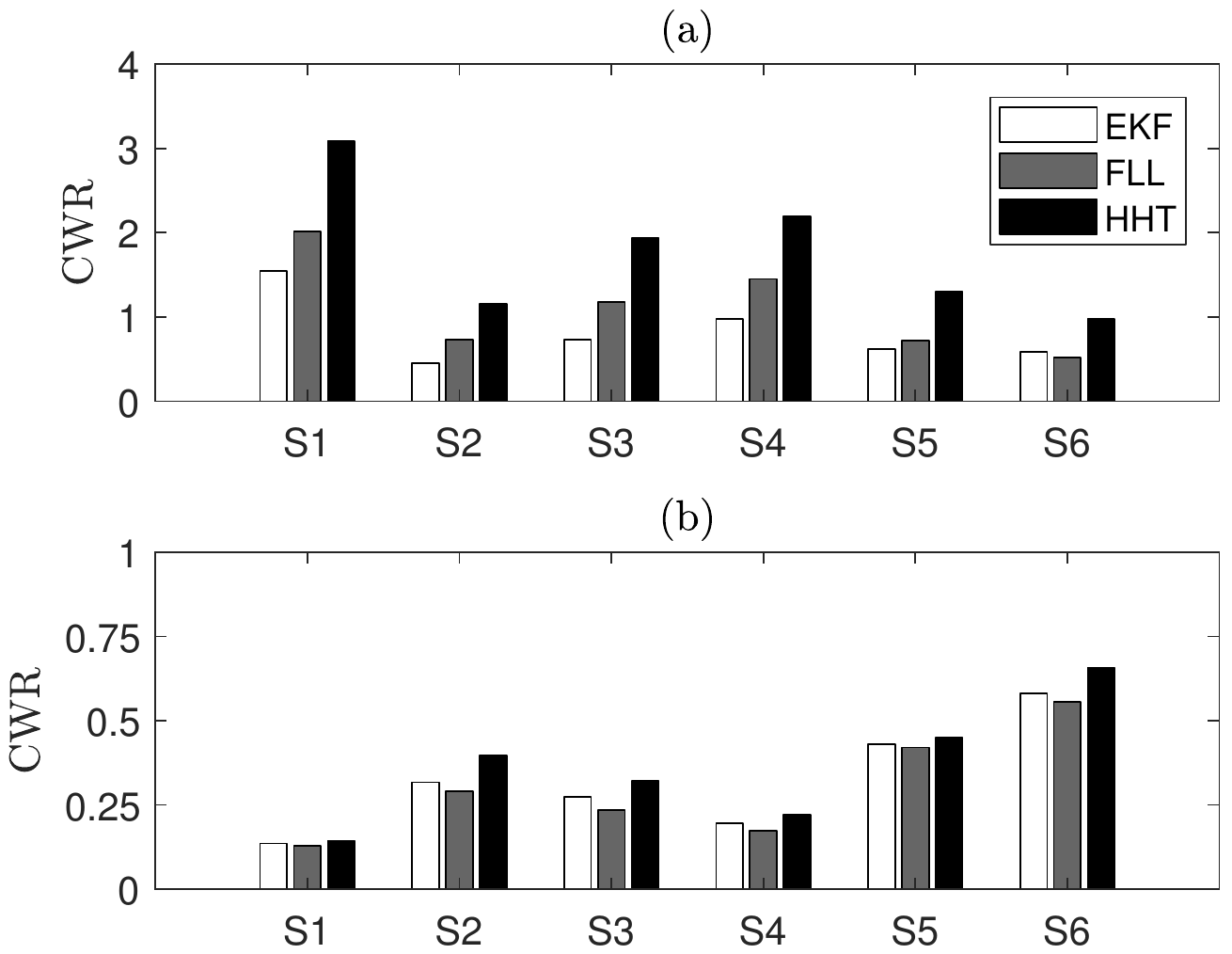}
\end{center}
  \caption{CWR under RC tuned at frequencies from the EKF, FLL and HHT (a) Unconstrained case (b) Constrained case.}
\label{fig:cwr_rc}
\end{figure}

\begin{figure}[!htpb]
\begin{center}
  \includegraphics[width=7.5cm]{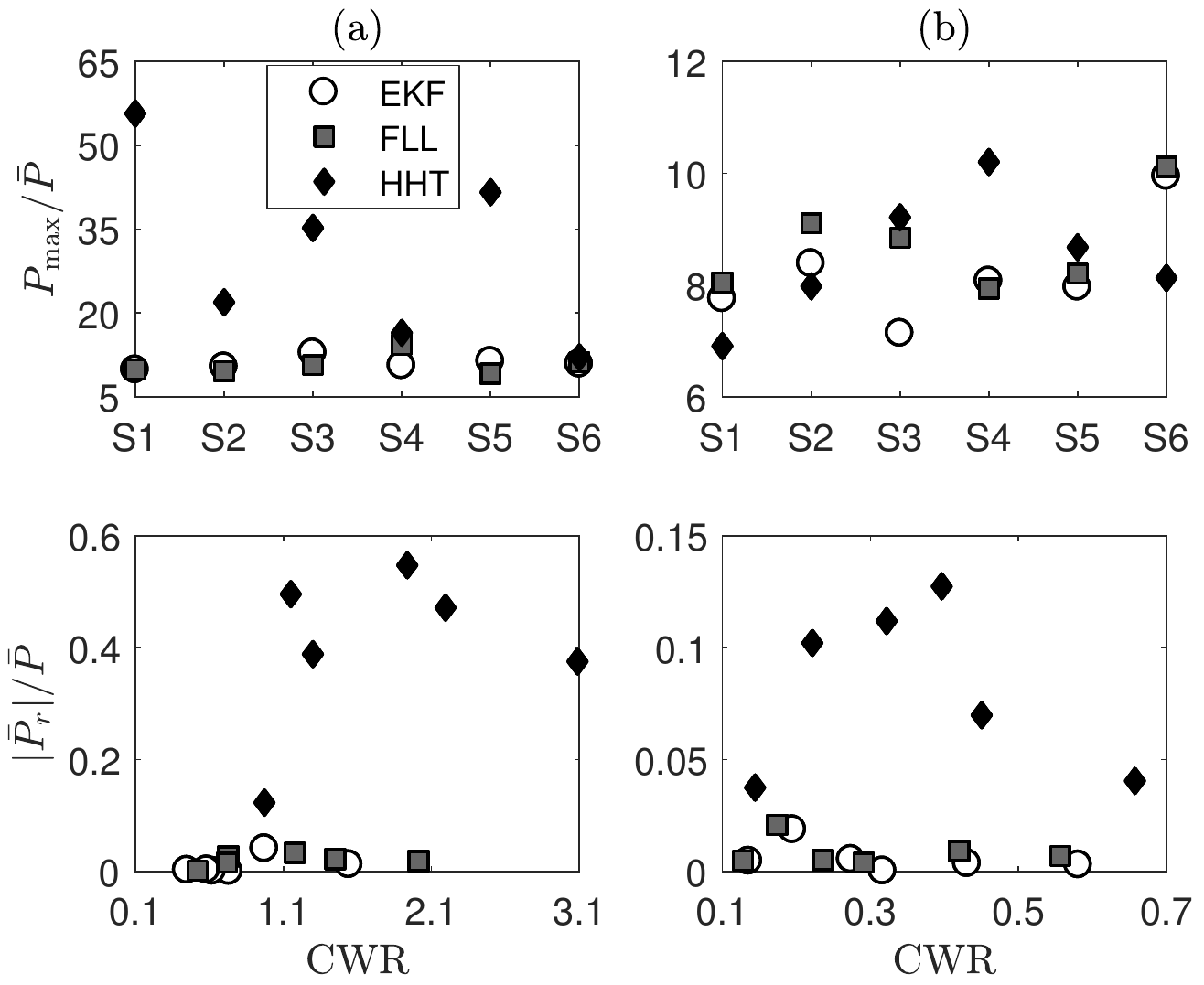}
\end{center}
  \caption{WEC performance under RC: Peak-to-average power ratio (top) and ratio of average reactive power and absorbed power as a function of CWR (bottom) (a) Unconstrained case (b) Constrained case.}
\label{fig:p2a_rc}
\end{figure}

Figures \ref{fig:cwr_rc} and \ref{fig:p2a_rc} illustrate the performance of the WEC, for the cases when the reactive control strategy adopts the EKF, FLL or HHT methods to estimate the wave excitation force frequency. For the unconstrained cases, the CWR is very large, in most situations, and the HHT obtains an energy improvement of up to $2.64$ over the EKF, or $1.67$ over the FLL. However, the body motion ranges from $-10$ to $10$\,m, which is practically impossible for a WEC with a draught of $4$\,m. Moreover, the PTO rating for the HHT approaches a factor of $50$ and the ratio of average reactive power and absorbed power is almost $60\%$ in some cases (Fig.\,\ref{fig:p2a_rc}.a). Such values would require oversized PTO equipment, which would not be a rational economic choice.

For the constrained cases, the improvement obtained from HHT over EKF or FLL is lower, but represents a more realistic scenario: an average energy improvement of $1.13$ is obtained when the HHT is compared to the EKF, or $1.23$ when compared to the FLL. Although the PTO rating varies from a factor of $7$ to $10$ for all frequency estimation methods, the ratio of average reactive power and absorbed power are much higher for the HHT, reaching almost $14\%$ (Fig.\,\ref{fig:p2a_rc}.b) for the studied cases.

\begin{figure}[!htpb]
\begin{center}
  \includegraphics[width=7.5cm]{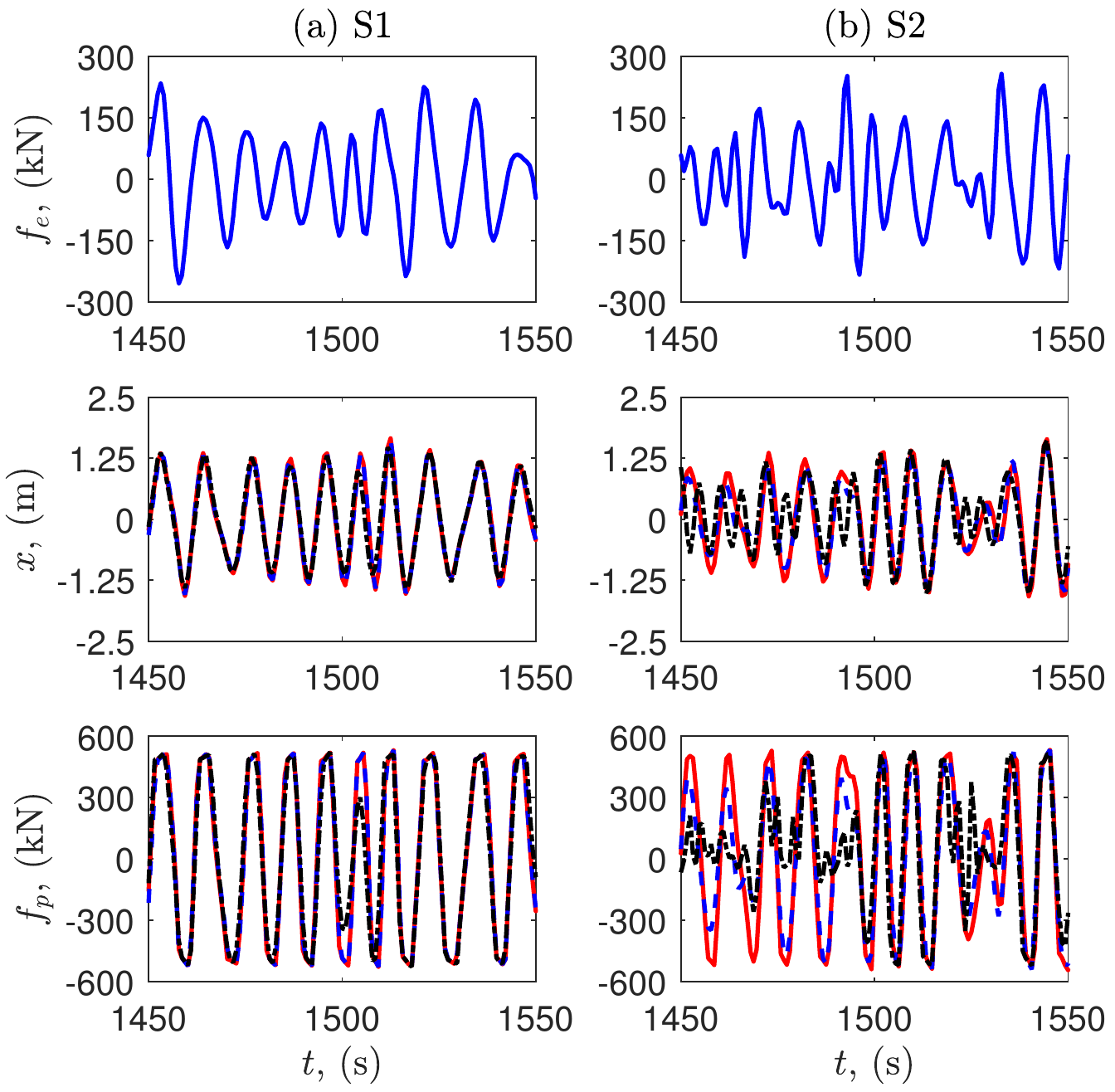}
\end{center}
  \caption{Time-series of the excitation force, position, and PTO force for the constrained RC tuned with the EKF (dashed blue line), the FLL (solid red line) and the HHT (dashed dotted black line) (a) S1; (b) S2.}
\label{fig:fe_x_fp_rc}
\end{figure}

\begin{figure}[!htpb]
\begin{center}
  \includegraphics[width=7.5cm]{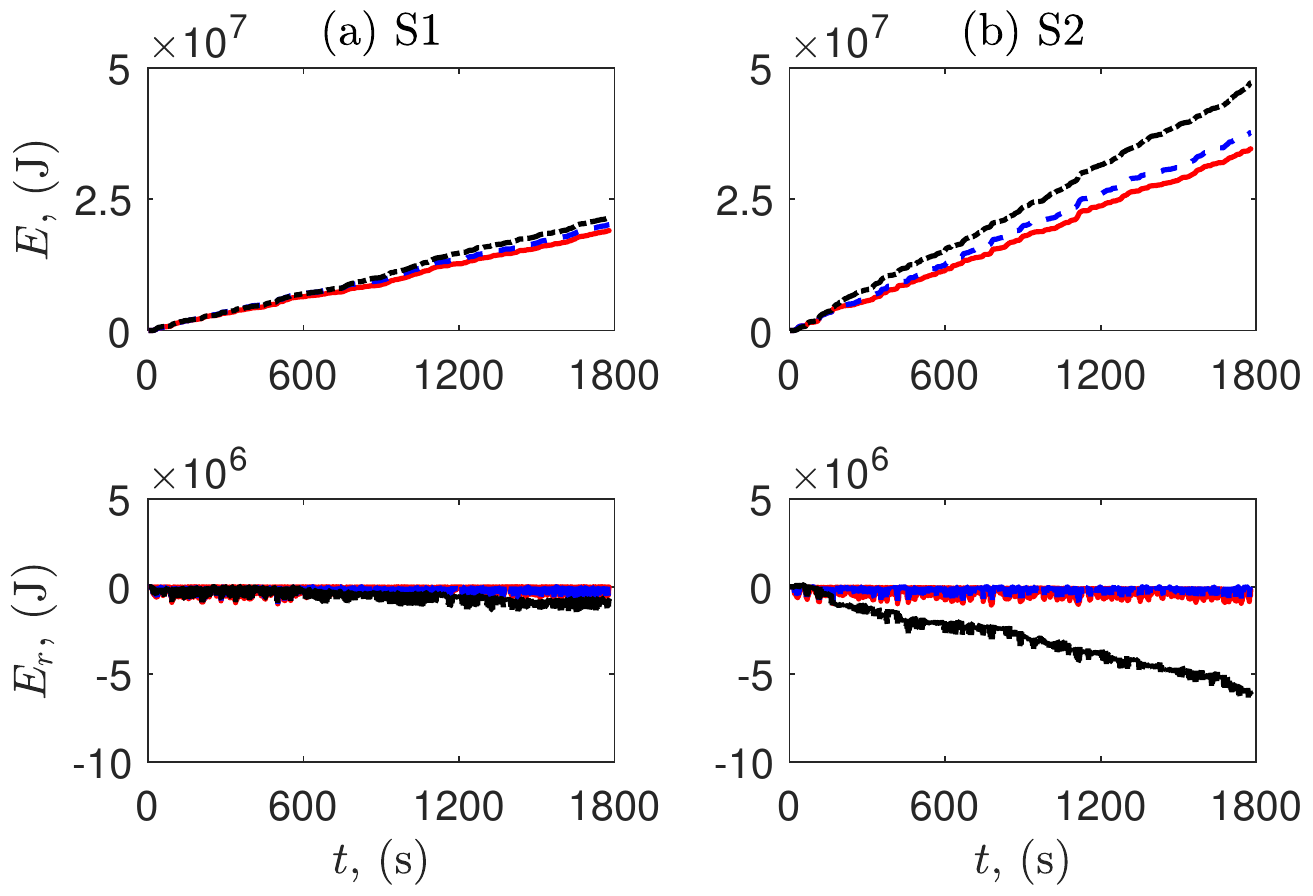}
\end{center}
  \caption{Energy absorbed (top) and reactive energy (bottom) over a 30-min simulation for the constrained RC tuned with the EKF (dashed blue line), the FLL (solid red line) and the HHT (dashed dotted black line) (a)S1; (b) S2.}
\label{fig:E_rc}
\end{figure}

In order to illustrate the effect of the frequency estimates on the variables of the system, for the constrained RC, Figure \ref{fig:fe_x_fp_rc} shows samples
of time-series simulation, and Figure \ref{fig:E_rc} shows the absorbed and the reactive energy over a 30-min simulation interval for sea states S1-S2. It can be noted that, for sea state S2, the reactive energy required for the RC, tuned with HHT frequency estimates is much higher than the FLL or EKF cases. In such a case, the HHT reactive power represents about $4.2\%$ of the total power, whereas the FLL reactive power is about $0.7\%$. Nevertheless, the absorbed power is $18\%$ greater with the HHT than with the FLL.


\subsection{Discussion}

In the EKF, a sinusoidal extrapolation method is used to model the excitation force as a monochromatic harmonic process, with varying amplitude and frequency. Thus, the EKF follows a single dominant frequency. Simulation results have shown that the estimated EKF frequency tracks the mean centroid frequency of the excitation force spectrum. The performance of the SOGI-FLL depends mainly on the appropriate selection of two design parameters: $\kappa$ and $\gamma$. Such parameters define the bandwidth of the adaptive filter and the FLL tracking frequency dynamics \cite{Rodriguez2012}. For the FLL, it has been shown that the mean estimated frequency converges to the energy frequency of the excitation force spectrum.

The HHT calculates the instantaneous frequency of the excitation force by decomposing the signal into a number of IMF components. Here, we have chosen only the IMF component with the highest energy content. This component has a wider bandwidth than the estimates from the EKF or FLL, and the CWR of the WEC is greater when the HHT frequency estimates are used for PC and RC. Even with a low energy content, the resonance frequency of the WEC is within the HHT estimates, which can also explain why tuning the controllers with these estimates result in the greatest energy absorption.


As expected, the PTO rating for PC is higher with the HHT than with the EKF, or FLL, in most of the cases. However, for the constrained RC, the greatest PTO rating does not indicate the greatest energy absorption. The average value of the PTO rating is about the same for all methods. Still, the greatest energy absorption, obtained with the HHT, also requires the greatest amount of reactive energy.

Here we have assumed the wave excitation force is known completely over the simulation interval. Both the EKF and FLL methods give an online estimate of the frequency, provided that an estimate of the excitation force is available. In the HHT method, we have adopted an off-line EMD algorithm, but a few implementation studies on the HHT have proposed real-time EMD algorithms, see, e.g., \cite{Hong2012}.


\section{Conclusion}
This paper has shown how different frequency estimation methods, used for controller tuning purposes, impact the energy absorbed by the WEC, the PTO rating and the required reactive power during the conversion process. The effect of the estimation methods on the WEC performance depend on the control strategy employed, the PTO system constraints, and the local wave spectrum.

For a control strategy that relies on the information of a dominant frequency component, such as the schemes in \cite{Fusco2013} and \cite{Tedeschi2012}, the EKF or FLL should be adopted. The mean frequency estimated by such methods converges, respectively, to the mean centroid frequency or the energy frequency of the spectrum. Moreover, if the sea state is characterized by a narrowband spectrum, the benefit of adopting a method that estimates the wave-to-wave frequency is relatively small.

By adopting a method that estimates the instantaneous frequency of the excitation force (the HHT), an average improvement in the energy absorbed of about 18\% is obtained over the EKF and FLL methods, for the constrained reactive control strategy. For passive control, an average improvement of 16\% is also obtained for the HHT. The greatest improvements of the HHT over the other methods are obtained for wideband spectra. In contrast to the EKF and FLL methods, where the bandwidth is narrow and the frequency estimates oscillate around a dominant frequency component, the HHT frequency estimates cover a wider range, and the location of the dominant frequency component is identified.


In this study, the first IMF component is adopted for the HHT approach. The frequency bandwidth of this IMF component could be narrowed by applying techniques that deal with mode mixing in EMD. Such an approach will be explored in future studies.

\ifCLASSOPTIONcaptionsoff
  \newpage
\fi


\bibliographystyle{IEEEtran}
\bibliography{pbgr2017}

\end{document}